\documentclass[aps,prx,twocolumn,superscriptaddress,showpacs,nofootinbib]{revtex4-1}

\usepackage{graphicx}
\usepackage{amsmath,color}
\usepackage{pgfplots}
\usepackage[export]{adjustbox}
\usepackage[normalem]{ulem}
\usepackage{float}

\def\gappeq{\mathrel{ \rlap{\raise.5ex\hbox{$>$}}
                      {\lower.5ex\hbox{$\sim$}} } }
\def\lappeq{\mathrel{ \rlap{\raise.5ex\hbox{$<$}}
                      {\lower.5ex\hbox{$\sim$}} } }
\usepackage{bm}
\usepackage{color}
\usepackage{epstopdf}

\newcommand\edd{\varepsilon_\text{dd}}
\newcommand{\del}[1]{\textcolor{red}{}}

\newcommand{\nick}[1]{\textcolor{black}{#1}}
\newcommand{\tom}[1]{\textcolor{black}{#1}}
\newcommand{\andrew}[1]{\textcolor{black}{#1}}

\begin{document}

\title{Quantum ferrofluid turbulence}

\author{T. Bland}
\author{G. W. Stagg}
\author{L. Galantucci}
\author{A. W. Baggaley}
\author{N. G. Parker}
\affiliation{Joint Quantum Centre Durham--Newcastle, School of Mathematics and Statistics, Newcastle University, Newcastle upon Tyne, NE1 7RU, United Kingdom}


\begin{abstract}
We study the elementary characteristics of turbulence in a quantum ferrofluid through the context of a dipolar Bose gas condensing from a highly non-equilibrium thermal state.  Our simulations reveal that the dipolar interactions drive the emergence of polarized turbulence and density corrugations.  The superfluid vortex lines and density fluctuations adopt a columnar or stratified configuration, depending on the sign of the dipolar interactions, 
\tom{with the vortices tending to form in the low density regions to minimize kinetic energy}.   
When the interactions are dominantly dipolar, \nick{the decay of vortex line length is enhanced,  closely following a $t^{-3/2}$ behaviour}.  This system poses exciting prospects for realizing stratified quantum turbulence and new levels of generating and controlling turbulence using magnetic fields.
\end{abstract}

\maketitle

In conventional ferrofluids, colloidal suspensions of permanently magnetized particles, the dipole-dipole inter-particle interaction gives rise to unique fluid properties, such as the normal field instability and flow characteristics which can be varied through an external magnetic field \cite{Rosensweig97}.  The ability to direct the fluid using magnetic fields has led to broad applications from tribology to targeted medicine \cite{Alexiou02}.  Remarkably, turbulence in ferrofluids has been limited to only a few studies \cite{Anton1990,Kamiyama1996,Schumacher2003}; this may be attributed to difficulties in achieving turbulent regimes (due to their high viscosity) and in characterising the flow (due to their opacity).  As such the manner in which the anisotropic, long-range interactions modify the turbulent state remains an open question.  Nonetheless, theoretical work has predicted that the coupling with ferrohydrodynamics leads to new turbulent phenomena such as control over the onset of turbulence through the applied magnetic field \cite{Altmeyer2015} and new modes of energy dissipation and conversion \cite{Schumacher2008}.

Quantum ferrofluids have been realised since 2005 through Bose-Einstein condensates of atoms with sizeable magnetic dipole moments - Cr \cite{griesmaier_2005,beaufils_2008}, Dy \cite{lu_2011,tang_2015} and Er \cite{aikawa_2012,chomaz_2016} - and have led to recent landmark demonstrations of self-trapped matter-wave droplets \cite{chomaz_2016,Schmitt2016,droplets} and the quantum Rosensweig instability \cite{kadau_2016,ferrier_2016}.  In combining ferrohydrodynamics with superfluidity, quantum ferrofluids embody a prototype system for studying ferrofluid turbulence due to the absence of viscosity and the quantization of vorticity.    As demonstrated experimentally for conventional condensates, states of such quantum turbulence can be formed and imaged \cite{Henn2009,Kwon2014,Navon2016}, and can show both direct analogies to its counterpart in everyday viscous fluids (for example, Kolmogorov scaling \cite{White2014,Tsatsos2016} and the transition from the von K\'{a}rm\'{a}n vortex street \cite{Kwon2016}) and distinct quantum effects (for example, ultra-quantum regimes \cite{White2014,Tsatsos2016} and non-classical velocity statistics \cite{White2010}), depending on the details of the turbulent state.  As well as the simplified fluid characteristics, these systems have the facet that the fluid parameters ({\it viz}. atomic interactions) can be tuned at will.  As such, quantum ferrofluids stand to shed light on general aspects of ferrofluid turbulence, as well as \nick{phenomena specific to the quantum nature of the fluid, such as quantized vortex line dynamics, reconnections and inviscid dissipation mechanisms} \andrew{\cite{Barenghi2014}}.


While various aspects of vortices in quantum ferrofluids have been theoretically explored, e.g. their generation, profiles and lattice structures \cite{Martin2017}, the behaviour of quantum ferrofluid turbulence remains at large.   Here we study turbulence in quantum ferrofluids through the scenario of a homogeneous dipolar Bose gas freely evolving from highly non-equilibrium conditions.  This scenario, representative of a sudden quench from a thermal gas through the BEC transition, is known to generate unstructured quantum turbulence which decays over time \cite{Berloff2002,Stagg2016}. 
 This setting, free from boundaries and artefacts that may be introduced  by external forcing, allows us to unambiguously identify the effects of the dipolar interactions.  

\begin{figure}
\includegraphics[width=1.05\columnwidth]{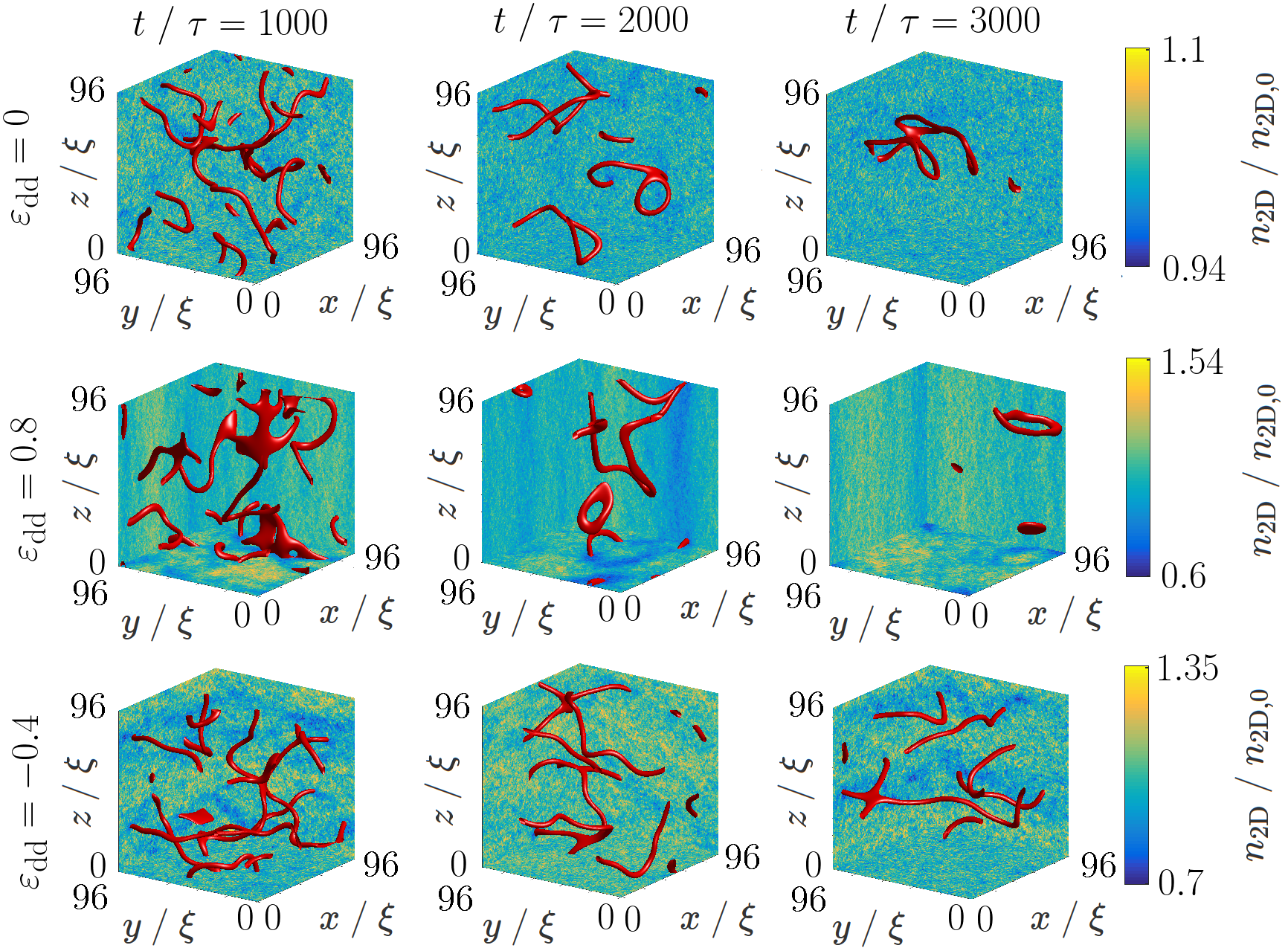}
\caption{Turbulence in the quantum ferrofluid, for three values of $\edd$. Vortices are shown through isosurfaces (red tubes) of the quasi-condensate density at the $0.05\langle n_q \rangle$ level (angled brackets denote ergodic average). The walls of the box show 2D density profiles, corresponding to integrating the \tom{density, $|\psi|^2$,} over the dimension perpendicular to their face.}
\label{fig:strat}
\end{figure}

We adopt the classical field methodology of the weakly-interacting, finite-temperature Bose gas \cite{Svis5,Davis,Sinatra2001,Berloff2002,Davis2,Connaughton2005,Pol_Rev,Proukakis,Blakie}, extending it to include dipolar interactions. 
The gas is described by a classical field $\psi({\bf r},t)$ (a valid assumption providing the modes are highly occupied), with atomic density $n({\bf r},t)=|\psi({\bf r},t)|^2$ and whose equation of motion is given by the Gross-Pitaevskii equation (GPE).    In the presence of dipolar interactions, the GPE given by \cite{Lahaye2009},
\begin{equation}
i \hbar \frac{\partial \psi}{\partial t}=\left [-\frac{\hbar^2}{2m}\nabla^2+g |\psi|^2+ \int U_{\rm dd}({\bf r}-{\bf r}')n({\bf r'},t)\,{\rm d}{\bf r}'\right ]\psi.
\label{eqn:dgpe1}
\end{equation}
The $g|\psi|^2$ term accounts for the local van der Waals-originating atomic interactions, parametrised by $g$.  The non-local integral term accounts for the long-range dipolar interactions, with interaction pseudo-potential $U_{\rm dd}(\mathbf{r-r'})=\dfrac{C_{\rm dd}}{4\pi}\dfrac{1-3\cos^{2}\theta}{|\mathbf{r-r'}|^3}$, with $\theta$ being the angle between the polarisation direction and the inter-atom vector  ${\bf r}-{\bf r'}$, and $C_{\rm dd}=\mu_0 d^2$, where $\mu_0$ is the permeability of free space and $d$ is the magnetic dipole moment of the atoms.  This potential accounts for the attraction of end-to-end dipoles and repulsion of side-by-side dipoles.  The relative strength of the dipolar interactions is specified by the ratio $\edd=C_{\mathrm{dd}}/3g$ \cite{Lahaye2009}.  \nick{Using well-established experimental techniques to tune $g$ via field-induced Feshbach resonances \cite{Lahaye2009} and the recent demonstration of tuning $C_{\rm dd}$ via fast rotation of the polarization direction \cite{lev}, the parameter $\edd$ can be experimentally varied over the range $-\infty \leq \edd \leq \infty$, including the regime of negative $C_{\rm dd}$ (in which side-by-side dipoles effectively attract and head-to-tail dipoles effectively repel).}

\nick{For $g>0$ and $-0.5 \leq \edd \leq 1$} the ground state of the dipolar Bose gas is the uniform solution $\psi=\sqrt{n_0}e^{iS_0}$, where $n_0$ is the uniform density and $S_0$ an arbitrary uniform phase, and chemical potential $\mu_0=n_0 g  \left(1- \varepsilon_{\mathrm{dd}} \right)$.  
According to Bogoliubov theory, perturbations to this state of momentum ${\bf p}$ have energy $E_{\rm B}(\mathbf{p})=\sqrt{c^2(\theta_k) \, p^2 + \left(p^2/2m \right)^2}$
where $c^2(\theta_k)=(gn_0/m) \left[ 1+ \varepsilon_{\mathrm{dd}}\left(3 \cos^{2} \theta_k-1 \right) \right]$ and $\theta_k$ is that between $\mathbf{p}$ and the polarization direction \cite{Lahaye2009}.  For small $p$ the spectrum corresponds to \nick{oscillatory excitations in the form of} phonons with anisotropic phase velocity  $c(\theta)$.  \nick{Note that outside of the regime of $g>0$, $-0.5 \leq \edd \leq 1$,} the excitations develop imaginary energy components, signifying unstable growth (the ``phonon instability") \nick{and the instability of this homogeneous state}.   

We express length in units of the dipolar healing length $\xi=\hbar/\sqrt{m \mu_0}$ and time in terms of the unit $\tau=\hbar/\mu_0$.
\tom{ The GPE is evolved numerically using a split step Fourier method \cite{Fleck1976,Fleck1978} on a $192^3$ periodic grid with spacing $d=0.5 \xi$.  Our findings are insensitive to these numerical parameters.  The time step $\Delta t =0.001 \tau$  is two orders of magnitude smaller than the timescale of the fastest modes supported \cite{modes}.} 
 Following previous approaches \cite{Berloff2002,Stagg2016} we initialize the system with the non-equilibrium state $\psi \left(\mathbf{r},0\right)=\sum_{\mathbf{k}}a_{\mathbf k}\exp(i\mathbf{k}\cdot\mathbf{r})$, where ${\bf k}$ is the wavevector (defined up to the maximum amplitude allowed by the numerical box \cite{Berloff2002,Connaughton2005}), \tom{the coefficients $a_\textbf{k}$ are uniformly valued (up to a certain wave vector amplitude set by the choice of system parameters), and the phases are distributed randomly \cite{ICs}.}  We illustrate the key behaviours through case studies of \tom{$\edd=0.8$ and $\edd=-0.4$}, as well as the non-dipolar case $\edd=0$ for comparison; the more general behaviour will also be described.

At very early times, there is a rapid self-ordering of the field, akin to the non-dipolar case \cite{Berloff2002,Connaughton2005}.  From the initially uniform distribution across modes, the low $k$ modes grow to develop macroscopic occupation, forming a quasi-condensate.  The high $k$ modes develop low occupations and are associated with thermal excitations.  Within of the order of 100 time units, this bimodal distribution across the modes has effectively saturated.  Unlike the non-dipolar case, the mode occupations are anisotropic in momentum space. The quasi-condensate has superfluid ordering and features a tangle of quantized vortices.  \tom{ To visualise the superfluid vortices for $t>0$ \cite{note}, we follow Ref. \cite{Berloff2002} in defining a ``quasi-condensate" density $n_{\rm q}$ of the low-lying modes $k \leq k_c$, where $k_c$ is identified from the condensate-thermal crossover in the mode distribution.  Here we identify $k_c = 0.46 \xi^{-1}$.  Vortices are then identified as tubes of low quasi-condensate density,  $n_q < 0.05 \langle n_q \rangle$, where $\langle \rangle$ denotes the ergodic average.   Our results are insensitive to the precise values of $k_c$ and the density threshold.}


In the non-dipolar Bose gas [Fig.~\ref{fig:strat} (top row)], this tangle is randomised in space, with no large-scale structure \cite{Berloff2002,Connaughton2005}, and the density fluctuations, representative of the high $k$ component of the field, are isotropic in space.   

For the dipolar Bose gas the spatial isotropy is broken. For $\edd>0$ [$\edd=0.8$, Fig.~\ref{fig:strat} (middle row)] the density fluctuations become columnar, aligned along the polarization direction, as seen in the integrated density profiles.  This is \nick{because these modes have lower energy, as seen in the earler dispersion relation $E_{\rm B}({\bf p})$, due to the lower energy configuration of dipoles to a head-to-tail} configuration. These fluctuations are sizeable in amplitude, ranging from around $0.6$ to $1.5$ of the mean density, and are dynamic. The vortices visibly tend to orient along $z$.

For $\edd<0$ ({\it viz.} $C_{\rm dd}<0$) [$\edd=-0.4$, Fig.~\ref{fig:strat} (bottom row)] the density fluctuations become planar, in accord with $E_{\rm B}({\bf p})$ and driven by the attraction of side-by-side dipoles, again with a large density amplitude.  The vortices in this case prefer to align in these low density planes.  

For all cases, the vortices decay in time, through reconnections, Kelvin wave decay and thermal dissipation, and by $t \approx 2000 \tau$ only a few vortex loops are left in the gas.
The columnar (planar) density fluctuations arise generically for $\edd>0$ ($\edd<0$), growing in amplitude with $|\edd|$.  
\tom{For larger $|\edd|$ values than shown, however, the dominance of the columnar (planar) density fluctuations makes it challenging to visualise the vortices.}

\begin{figure}
\includegraphics[scale=0.9]{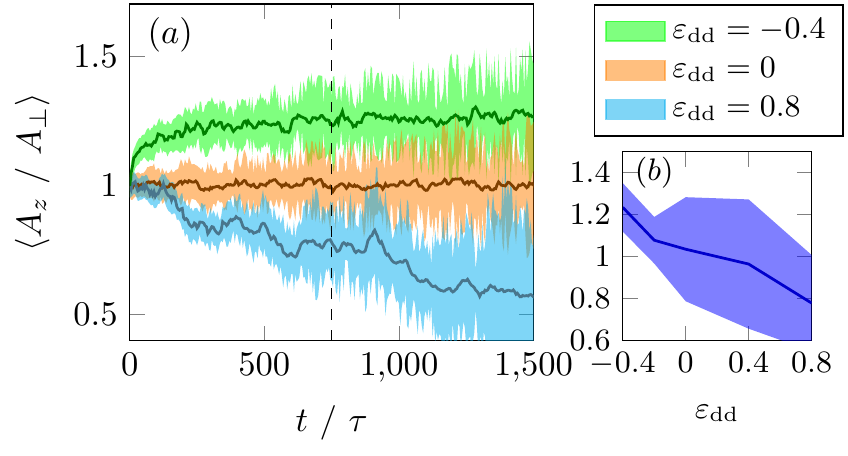}
\caption{$(a)$ Polarization of the vortex tangle over time, shown through the area ratio of vortices, $A_z/A_\perp$,  for three $\edd$ values.  Lines and shaded regions represent the mean and one standard deviation over five realizations with different randomised initial conditions. $(b)$ Snapshot of the area ratio over more detailed range of $\edd$ at $t=750\tau$ (dotted line in $(a)$).  }
\label{fig:area}
\end{figure}

Next we quantify the polarization of the vortex tangle.  We project the quasi-condensate vortex tubes in the $x$, $y$ and $z$ directions, denoting the areas cast as $A_x$, $A_y$ and $A_z$, respectively.    The ratio $A_z/A_\perp$, where $A_\perp=\frac12(A_x+A_y)$, then quantifies the axial-to-perpendicular anisotropy of the vortices.   Figure~\ref{fig:area}$(a)$ shows the evolution of $A_z/A_\perp$, over five initial conditions for each $\edd$, \nick{up to $t=1500\tau$.  By this time, the number of vortices has decreases to the order of unity; beyond this the area ratio is no longer a meaningful characteristic of the tangle, with the fluctuations becoming excessive}.  
\tom{Moreover, at this stage the dynamics are no longer mutli-scaled, a key characteristic of hydrodynamic turbulence.}
Due to the isotropic initial conditions, all cases begin being isotropic with $A_z/A_\perp \approx 1$.
For $\edd=0$, the tangle remains isotropic throughout.  However, for $\edd\neq 0$ the tangle evidently becomes polarized, seen by the statistically significant deviation of $A_z/A_\perp$ from unity.  For $\edd=0.8$, $A_z/A_\perp$ decreases by up to 25\%; for $\edd=-0.4$ it increases up by $40\%$.  In Fig.~\ref{fig:area}$(b)$ a more thorough parameter sweep of $\edd$ is displayed, focussing on the asymptotic value of $A_z / A_\perp$ obtained at $t / \tau=750$, where, within errorbars, this quantity decreases approximately with $\edd$.
 
\begin{figure}
\includegraphics[scale=0.8]{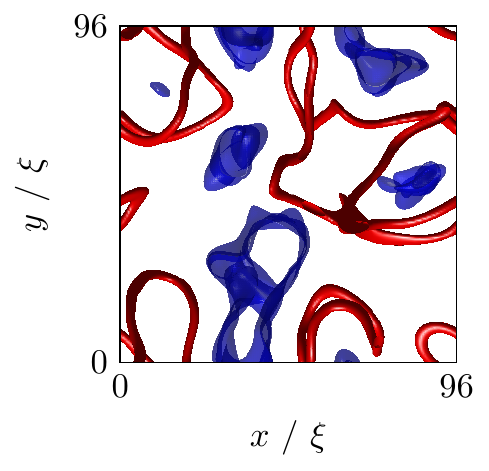}
\raisebox{0.6cm}{\includegraphics[scale=0.3]{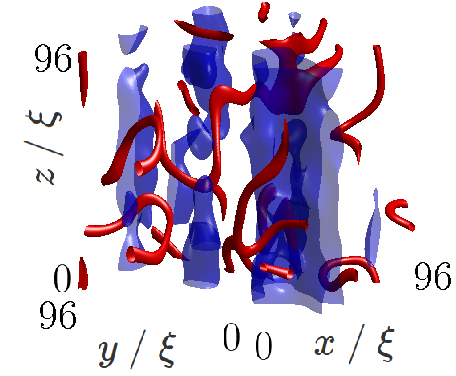}}
\caption{A representative snapshot (left: top view; right: angled view) for $\edd=0.8$ ($t / \tau=1000$) highlighting the location of the high density regions (blue isosurfaces, plotted at $0.8\,\text{max}(\langle n_q\rangle)$) and the vortices (red isosurfaces, plotted at 0.05$\langle n_q \rangle$).  }
\label{fig:droplets}
\end{figure}

\nick{To further understand the polarization of the vortices, Fig. \ref{fig:droplets} shows the location of the vortex lines (red isosurface tubes) and regions of high quasi-condensate density (blue isosurface regions).  It is evident that, firstly, the system is threaded with vertical tube-like structures of high density; the intervening regions being of low density.  Secondly, the vortices avoid the high density regions; by maximising their overlap with the low density regions they reduce their kinetic energy.  For $\edd<0$ the high density regions are planar strata, with the vortices tending to locate in the intervening low density layers.}

\begin{figure}
\includegraphics[scale=0.35]{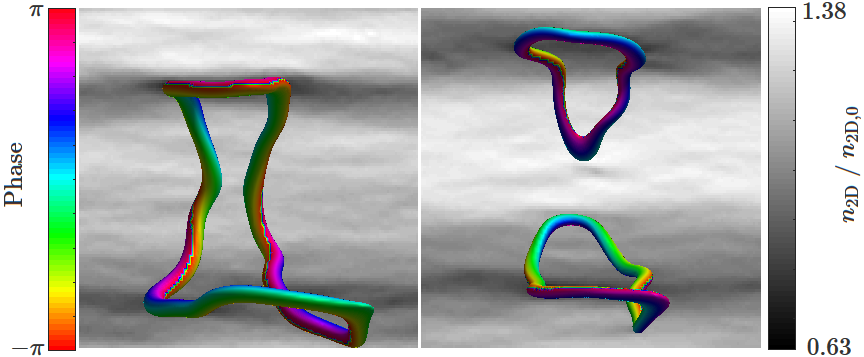}\\
\caption{Coupling between vortices and the dipolar-driven density fluctuations.  One large loop spreads across two planar density regions, which act as pinning layers. \tom{The observed dynamics are indicative of simulations for relative dipolar strength $\edd<0$ and the times of the images shown are $t / \tau=50$ apart from left to right}. The phase is \tom{evaluated and displayed on} the quasi-condensate density at the $0.05\langle n_q\rangle$ level, the back wall of the box shows the 2D density profiles, corresponding to integrating the density over the dimension perpendicular to its face.}
\label{fig:planar_vt}
\end{figure}

\nick{The preference of the vortices to align in the low density regions is particularly prominent at the late stages of the decay, when only one or a few vortex loops remain.  Here we observe situations, for example, in which vortex loops becomes heavily pinned across two planar regions of low density, as in  Fig.~\ref{fig:planar_vt}.}  Considerable vortex line length lies in these planes, while two vortex segments connect between these planes to form the overall loop.   The pinned segments move with the low density region.  This large loop is metastable but decays eventually via a reconnection, forming two small loops, each of which is heavily pinned within each low density plane.  We observe such pinning of vortices to the low density region to be a general occurrence for moderate to large values of $\edd$.

\begin{figure}
\includegraphics[scale=0.85]{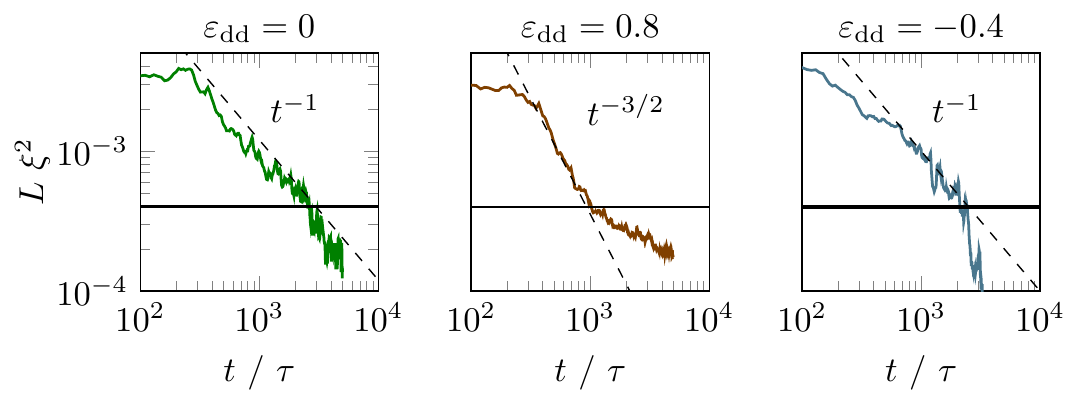}
\caption{Decay of the vortex tangles.  Vortex line length $L$ for the three $\edd$ values shown, averaged over five simulations.  For reference, $t^{-1}$ and $t^{-3/2}$ lines are shown (dashed lines) where appropriate.  \nick{The horizontal line shows when the average intervortex spacing $\ell=1/\sqrt{L}$ equals the box half-width, representing the level at which the number of vortices becomes of the order of unity.}}
\label{fig:line_length}
\end{figure}

Having identified an additional `organization' of the vortices driven by the dipolar interactions we now seek to understand how this affects the nature of the turbulence itself.
In contrast with classical turbulence, two distinct regimes of quantum turbulence have been identified \cite{Walmsley2008}. In the quasi-classical regime, motions over a wide range of scales are observed, and many of the statistical properties of classical turbulence (such as Kolmogorov's energy spectrum) are observed \cite{Baggaley-Laurie-2012}.  
However quantum fluids also give rise to another form of turbulence, the ultraquantum or Vinen regime, which is associated with a random tangle of quantized vortices and no large-scale structure.  In the quasi-classical regime energy dissipation is dictated by the lifetime of the largest scales of motion, and one can construct a classical argument that the rate of the decay of the vortex line density, $L$,  follows a power law scaling $L\sim t^{-3/2}$. The Vinen regime can be distinguished from the quasi-classical regime because different dissipation mechanisms dominate its decay, which leads to a different power-law scaling, with $L$ decaying as $t^{-1}$.

The turbulence arising from a thermally quenched (non-dipolar) Bose gas has been linked to ultra-quantum turbulence \cite{Stagg2016}.  Here we estimate the vortex line length $L$ as the volume occupied by the vortex tubes divided by their typical cross-sectional area \cite{Stagg2016,lines}; its evolution is shown in Fig.~\ref{fig:line_length}.  \nick{We interpret the behaviour above the horizontal line; below this, the number of vortices becomes of the order of unity}. For $\edd=0$ we recover the $t^{-1}$ behaviour of ultra-quantum turbulence.  \nick{The dynamics for $\edd=-0.4$ also closely follow this trend.} However, what is particularly striking is that for $\edd=0.8$ we see a faster decay, akin to $t^{-3/2}$.  \nick{At first glance, this suggests that the turbulence enters the quasi-classical regime.  To test this, we examine the lengthscale of the velocity correlations.}

\tom{We first calculate the longitudinal velocity correlation function $f_j(r,t)=\langle v_j({\bf x},t)v_j({\bf x}+r\hat{\bf e}_j,t)\rangle/\langle v_j({\bf x},t)^2\rangle$ along
each direction $j = x, y, z$, where $\bf v$ is the quasi-condensate velocity and the ensemble average is performed over positions $\bf x$.}
\nick{From this, we calculate the integral lengthscale $I_j(t)=\int_0^\infty f_j\, dr$, a convenient measure of the distance over which velocities are correlated \cite{Davidson2004,Stagg2016}.  For all cases of $\edd$, the integral lengthscales are significantly less than the average distance between vortices \tom{$\ell=1/\sqrt{L}$}: this confirms that there are no large-scale motions and that the turbulence is of the ultraquantum/Vinen form.  Furthermore, for $\edd=0$ and $\edd=-0.4$, we find that the integral lengthscale is isotropic ($I_x\approx l_y \approx l_z$) to within statistically fluctuations.  However, for $\edd=0.8$ we find that $I_z \approx 2 I_x \approx 2 I_y$, indicative of significant extension of the velocity correlations along the polarization direction.  This strong anisotropy in the velocity correlations may be responsible for the $t^{-3/2}$ scaling; however, a dimensional derivation of this situation which confirms this decay law remains outstanding.}




To summarise, for the first time we have numerically studied turbulence in a quantum ferrofluid. 
In the absence of dipolar interactions the rapid quench of a thermal gas through the transition temperature generates a random unstructured tangle with no significant large scale motions, that is, ultraquantum/Vinen turbulence. 
We find that for values of $\edd$ approaching unity, where the dipolar atomic interaction is comparable to the isotropic van der Waals interactions, the quantum turbulence that emerges is strongly polarized, \nick{both in the orientation of the vortex lines and the velocity correlations of the flow.  Whereas polarized quantum turbulence has been predicted in rotating superfluids \cite{Tsubota2004}, here the origin is very different, arising naturally from the inter-particle interactions without external forcing.}  In contrast for large negative values of $\edd$  the vortices arrange into sheets; this has the potential to lead to stratified quantum turbulence, which as yet is unexplored.

We believe that turbulence in a quantum ferrofluid will allow both experimental and theoretical studies of new and interesting aspects of fluid dynamics. For example the inverse cascade has received  much attention in quantum fluids recently \cite{Reeves_2013}, and it is entirely conceivable that new regimes of two dimensional turbulence can be realised by the presence of dipolar interactions within the gas. Finally whilst numerous mechanisms for continuously forcing three-dimensional turbulence in a BEC have been put forward \cite{Kobayashi,Henn2009,Navon2016}, most follow James Bond's lead and shake, rather than stir the condensate, generating significant phonon excitations \cite{Navon_2016}. By using a time dependent external magnetic field, or changing the effective value of $\edd$ (through modulation of the local van der Waals force $g$ for example) in both space and time one could stir the fluid in a method analogous to the magnetic stirring of a classical electrically conducting fluid \cite{Tabeling_1991}.

Data supporting this publication is openly available under an Open Data Commons Open Database License \cite{data}.

{\it Acknowledgements} - TB and NGP thank the Engineering and Physical Sciences Research Council (Grant No. EP/M005127/1) for support.


\begin{thebibliography}{10}
\bibitem{note} \nick{The identification of the quasi-condensate vortices is meaningful only for $t>0$.  At $t=0$ there is no quasi-condensate.}
\bibitem{ICs} \nick{While this isotropic initial condition is an approximation to the (anisotropic) pure thermal state of the dipolar gas, the system very rapidly evolves to an anisotropic thermal state.  This occurs on much shorter timescales than the vortex dynamics of interest.}
\bibitem{lines} \nick{We have confirmed that this method of evaluating the vortex line length gives comparable results (within 10\%) of a more accurate method based on tracking lines of pseudo-vorticity \cite{villois}.  }
\end{thebibliography}
\end{document}